\documentclass[twocolumn,aps,pre,showpacs,floatfix,preprintnumbers,superscriptaddress]{revtex4-1}
\usepackage{graphicx}
\usepackage[]{subcaption}
\usepackage{amsmath}
\usepackage{amssymb}
\usepackage{mathrsfs}
\usepackage{ulem,soul}
\usepackage{xcolor}

\usepackage[utf8]{inputenc}

\DeclareMathAlphabet{\mathpzc}{OT1}{pzc}{m}{it}
 
\begin{document}
\preprint{PI/UAN-2020-679FT}

\title{About the Teaching of Plane Motion of Rigid Bodies}

\author{Diego Luis González}
\email{diego.luis.gonzalez@correounivalle.edu.co}
\affiliation{Departamento de Física, Universidad del Valle, A.A. 25360, Cali, Colombia}

\author{Alejandro Gomez Cadavid}
\email{alejandro.cadavid@correounivalle.edu.co}
\affiliation{Departamento de Física, Universidad del Valle, A.A. 25360, Cali, Colombia}

\author{Yeinzon Rodr\'iguez}
\email{yeinzon.rodriguez@uan.edu.co}
\affiliation{Centro  de  Investigaciones  en  Ciencias  Básicas  y  Aplicadas,  Universidad  Antonio  Nariño, Cra  3  Este  \#  47A-15,  Bogotá D.C.  110231,  Colombia}
\affiliation{Escuela  de  F\'isica,  Universidad  Industrial  de  Santander, Ciudad  Universitaria, Bucaramanga  680002,  Colombia}  
\affiliation{Simons  Associate  at  The  Abdus  Salam  International  Centre  for  Theoretical  Physics, Strada Costiera  11,  I-34151,  Trieste,  Italy}

\date{\today}

\begin{abstract}
The study of the motion of a rigid body on a plane (RBP motion) is usually one of the most challenging topics that students face in introductory physics courses. In this paper, we discuss a couple of problems which are typically used in basic physics courses, in order to highlight some aspects related to RBP motion which are not usually well understood by physics students. The first problem is a pendulum composed of a rod and disk. The angular frequency of the pendulum is calculated in two situations: disk fixed to the rod and disk free to spin. A detailed explanation of the change in the angular frequency from one case to another is given. The second problem is a ladder which slides touching a frictionless surface. We use this problem to highlight the fact that the contact forces applied by the surface perform translational and rotational work despite that the total mechanical energy of the ladder is conserved. 
\end{abstract}



\maketitle

\section{Introduction}\label{sec:intro}

Two basic objectives that students must achieve in learning physics are the understanding of fundamental concepts, as well as the relationship between them. A basic diagnostic tool to measure the understanding level of the students is their performance applying basic concepts to solve physical problems. However, in a typical physics course, there are often many students who have difficulty to reach these objectives. For example, it is very common the misconception that there will always be a force in the direction of motion. Therefore, it is not a surprise that just a fraction of the students realize that the direction of motion is given by the velocity and not by the net force. One of the most challenging topic in introductory physics courses is the motion of a rigid body  on a plane (RBP motion). In the RBP motion, all the particles belonging to the body move in fixed planes which are all parallel to the frame of reference i.e., the motion is bidimensional. Consequently, the equations related to the RBP motion can involve up to three coordinates, two for the position of the center of mass ($cm$) and, an additional one, for the angular position of the body. Given the complexity of the RBP motion, several theoretical aspects related to the teaching of it have been treated in previous papers \cite{lopez,Salazar,Carnero,Hierrezuelo,Pinto,Sharma}. The learning difficulties and typical misconceptions of the students, associated to the RBP motion, have been reported previously by several authors \cite{lopez,Phommarach,Ortiz,Rimoldini,unsal,Mashood,Close,Khasanah,Rahmawati,Carnero,graham,Fang}. For instance,  the students have difficulty in determining the magnitude of the torque and connecting the total torque to the direction of rotation \cite{Gina}. It has been also reported that the students typically are in trouble at understanding that, in general, different points of a rigid body have different velocities and accelerations \cite{Gray}.

\begin{figure*}[htp]

\begin{subfigure}[b]{1\textwidth}
\begin{subfigure}{0.23\linewidth}
\includegraphics[scale=0.3]{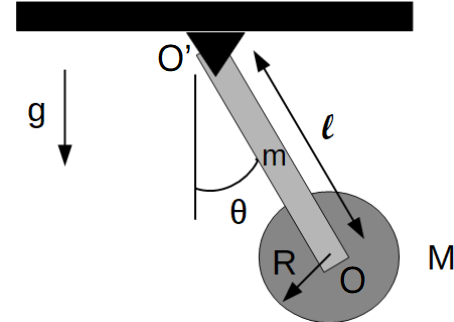}
\caption{}
\end{subfigure}
\begin{subfigure}{0.23\linewidth}
\includegraphics[scale=0.3]{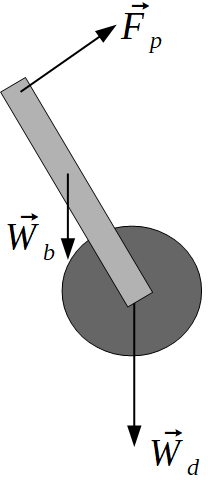}
\caption{}
\end{subfigure}
\begin{subfigure}{0.23\linewidth}
\includegraphics[scale=0.3]{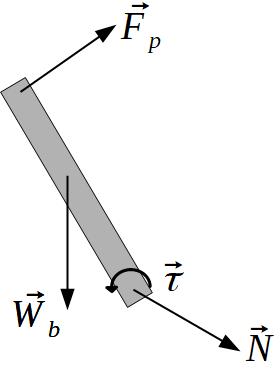}
\caption{}
\end{subfigure}
\begin{subfigure}{0.23\linewidth}
\includegraphics[scale=0.3]{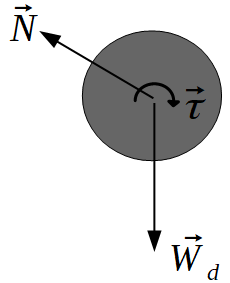}
\caption{}
\end{subfigure}
\end{subfigure}

\caption{(a) Pendulum composed by a disk and a rod. (b) External forces acting on the rod-disk rigid system. In (c) and (d) the forces acting on the disk and rod are shown, respectively. The interaction between the disk and rod is represented by the force $\vec{N}$ and the torque $\vec{\tau}$.}

\label{dr}
\end{figure*}

\begin{figure*}[htb]

\begin{subfigure}[b]{1\textwidth}
\begin{subfigure}{0.4\linewidth}
\includegraphics[scale=0.45]{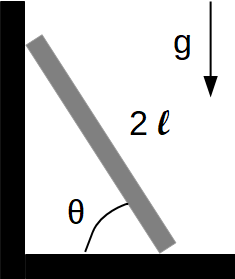}
\caption{}
\end{subfigure}
\begin{subfigure}{0.4\linewidth}
\includegraphics[scale=0.45]{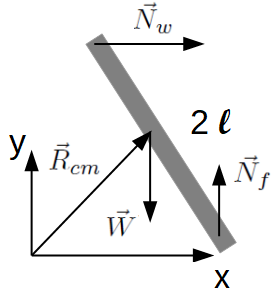}
\caption{}
\end{subfigure}
\end{subfigure}

\caption{(a) A ladder of length $2\ell$ starts to slide from rest on a frictionless surface and (b) forces acting on the ladder.}
\label{ladder}
\end{figure*}

In this paper we focus in two situations related to the RBP motion which usually are not well understood by the students of \textit{introductory} physics courses, that is why we restrict to the physical and mathematical tools given in this kind of courses instead of using more advanced tools such as the Lagrangian formalism. 

The first situation is related to the contribution of the moment of inertia in the angular frequency of a body composed by several rigid parts. Particularly, we consider a compound physical pendulum, which consists of a disk placed at the end of a rod as shown in Fig. \ref{dr} (a). In Ref. \cite{kleppner} this system is proposed as a problem that challenges the students to determine the oscillation frequency of the system in two cases. In the first case, the disk is fixed to the rod forming a single rigid body. In the second, the disk is mounted to the rod by a frictionless bearing so that it is perfectly free to spin. This exercise was chosen because it is very illustrative for our purpose and surprisingly, a quick search in informal physical forums shows that many of the reported solutions do not clearly explain why the oscillation frequency changes from one case to the other even though when it is a common problem that can be found in different course materials \cite{sol1,sol2,sol3,sol4}. Thus, our purpose here is to show that the students must realize how the angular momentum is divided into two parts and its implications in the oscillation frequency.




On the other hand, there are two basic approaches to write the work-energy theorem in the RBP motion. In the first one,  the system is considered as a collection of particles and the total work over the body, $\mathcal{W}$, is the sum over the work of each particle belonging to the system. Therefore, $\mathcal{W}$ is given by:
\begin{equation}\label{wsum}
\mathcal{W}=\sum_i\int_{\mathcal{C}_i} \vec{F}^{ext}_i\cdot d\vec{r}_i=\Delta E_k,
\end{equation}
where $\vec{F}^{net}_i$ is the total force acting on the $i$-th particle which is the sum of the forces applied by other particles belonging to the rigid body ($\vec{F}^{int}_i$) plus forces applied by particles that are not part of the rigid body ($\vec{F}^{ext}_i$) \cite{kleppner,spiegel}. In the second approach for the work-energy theorem, the motion of each particle is decomposed in translation of the center of mass ($cm$) and rotation about the $cm$. Consequently, in this last approach the total work over the system has two different contributions, one from the translation of the $cm$, $\mathcal{W}_T$, and another one from rotation, $\mathcal{W}_{\theta}$.

The second situation we want to discuss is related to the contributions of the external forces to the translational and rotational works. In several cases, the total mechanical energy of a body is conserved, although there are forces which perform translational and rotational work. Particularly, we consider another typical example: a ladder that slides without friction as in Fig. \ref{ladder} (a) \cite{kleppner,spiegel,meriam,lawden,sol5,sol6}. The forces associated to the wall and floor perform rotational and translational works which have the same magnitude but opposite sign. The students often tend to assume that the translational and rotational works associated to these forces are zero instead of realizing that the joint contribution to the work, $\mathcal{W}_{\theta}+\mathcal{W}_T$, is the one which vanishes. 

We choose to discuss the composed pendulum and the sliding ladder because they are classical examples of RBP motion often used in introductory physics courses to explain several concepts such as torque, angular momentum, energy conservation, etc. Additionally, they allow us to illustrate our objectives, so that the solutions presented in this paper are more detailed than those presented in \cite{sol1,sol2,sol3,sol4,sol5,sol6}. For example, in the ladder problem we use the integral of the work-energy theorem instead of writing the energy conservation equation.

This paper is divided as follows. In Sec. II we present a brief description of the basic equations of RBP motion. In Sec. III and IV we apply these equations to the two systems previously mentioned: the composed pendulum and the sliding ladder. Finally, in Sec. V we provide a discussion of the results and give some advices to improve the understanding of the students about the RBP motion.

\section{Plane Rigid Body Dynamics}

A body with mass $M$ can be considered as a rigid body if, for any couple of points $p$ and $p'$ of the body, the condition: 
\begin{equation}\label{rbdefn}
||\vec{r}_{p}-\vec{r}_{p'}||=C_{pp'},
\end{equation}
is satisfied with $C_{pp'}$ being a constant and $\vec{r}_p$, $\vec{r}_{p'}$ being the positions of the points. Therefore, for a rigid body, the relative distance between any couple of points belonging to the body remains constant.

There exist two simple ways in which a rigid body whose motion is restricted to a plane can move, it can translate without rotating or rotate about a fixed axis $O'$ without translating. In the first one, the motion of the body is fully specified by the translational dynamics of the $cm$. Consequently, the body motion is completely described  by the Newton's second law, i.e., the net force over the body, $\vec{F}_{net}$, is proportional to the acceleration of the $cm$, whose position, $\vec{R}_{cm}$, is defined by the relation  $M\vec{R}_{cm}=\int_{\Omega}\,dm \,\vec{r}$, where the integral extends over the body region $\Omega$.

The translational work done by the net force moving the $cm$ along the path ${\cal{C}}$ is given by:

\begin{equation}\label{workcmn}
\mathcal{W}_T=\int_{\mathcal{C}}d\vec{R}_{cm}\cdot \vec{F}_{net} = \Delta E_K,
\end{equation}
where $E_K= M\,V_{cm}^2/2$ and $\vec{V}_{cm}$ are the translational kinetic energy and the velocity of the $cm$, respectively. It is important to emphasize that Eq. (\ref{workcmn}) is completely general and still valid even if the body rotates on the plane.

In the second simple case, the axis may be at rest or moving with constant velocity without any change in its orientation, then it is convenient to put the origin of the inertial frame on the rotation axis $O'$. Besides, the velocity of an arbitrary point of the body with position $\vec{r}$ measured from $O'$ can be written as $\vec{v}=\vec{\omega} \times \vec{r}'$ where $\vec{\omega}$ is the angular velocity of the rigid body. In general, $\vec{v}$ is not equal for the different points of the body but $\vec{\omega}$ is the same for all of them. The rotational work $\mathcal{W}_{\theta}$ is defined by means of the net torque over the body, which is proportional to the angular acceleration $\vec{\alpha}$.

\begin{equation}\label{rotfan}
\mathcal{W}_{\theta} = \int_{\theta_A}^{\theta_B}   \! \vec{\tau}_{net} \cdot d\vec{\theta} = \Delta E_K,
\end{equation}
where $E_K= I_{O'}\,\omega^2/2$ is the rotational kinetic energy and $I_{O'}=\int_{\Omega} dm\,r^2$ is the moment of inertia about $O'$. It is important to emphasize that, for a rigid body, $I_{O'}$ just depends on the mass distribution relative to $O'$. Thus,  it is clear that $I_O'$ is independent on the motion state of the body.

\begin{figure}[htp]
\begin{center}
\includegraphics[scale=0.5]{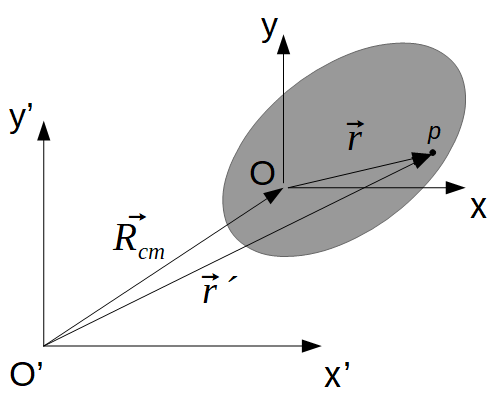}
\end{center}
\caption{$\vec{R}_{cm}$ and $\vec{r}'$ represent the positions of the $cm$ and of the point $p$ relative to the inertial reference system $O$, respectively. Analogously, $\vec{r}$ is the position of $p$ relative to the $cm$. }
\label{cman}
\end{figure}

The most general case of RBP motion is given when the rotation axis $O$ moves on the plane.
In this case, it is convenient to decompose the motion of the body in displacement of the $cm$ and rotation about the $cm$. This decomposition can be archived noticing that the position of any point of the body, $\vec{r}\,'$, can be written as the sum of the position of the center of mass $\vec{R}_{cm}$ and the relative position of the point with respect to to the $cm$, $\vec{r}$, as shown in Fig. \ref{cman}.  Thus, by using the decomposition $\vec{r}\,'=\vec{r}+\vec{R}_{cm}$, the angular momentum is given by:

\begin{eqnarray}\label{lrn}
\vec{L}&=&\int_{\Omega} \vec{r}\hspace{0.01cm}^{'} \times d\vec{p}\,^{'}=\int_{\Omega} dm\,(\vec{r}+\vec{R}_{cm})\times\frac{d}{dt}(\vec{r}+\vec{R}_{cm})\nonumber\\
&=&\vec{R}_{cm}\times \vec{P}_{cm}+I_{cm}\,\vec{\omega},
\end{eqnarray}
where $I_{cm}$ is the moment of inertia relative to the $cm$ and $\vec{P}_{cm}$ the linear momentum of the center of mass. Hence, the net torque can be found by taking the time derivative on Eq. (\ref{lrn}):
\begin{equation}\label{tau1n}
\vec{\tau}_{net}=\vec{R}_{cm}\times\vec{F}_{net}+I_{cm}\vec{\alpha}.
\end{equation}
If the body describes a pure rotation about the fixed axis $O'$, then $\vec{V}_{cm}=\vec{R}_{cm}\times\vec{\omega}$ and Eqs. (\ref{lrn}) and (\ref{tau1n}) take the form:

\begin{equation}\label{n1n}
\vec{L} = I_{O'}\vec{\omega}\,\,\,\,\text{and}\,\,\,\,\vec{\tau}_{net}=I_{O'}\vec{\alpha},
\end{equation}
where $I_{O'}$ is the moment of inertia about $O'$. 

Let $d\vec{F}^{ext}(\vec{r})$ be a differential of the total external force over the body which is applied at position $\vec{r}$, and, $\vec{F}_{net}=\int_{\Omega} d\vec{F}^{ext}(\vec{r})$. Consequently $\vec{\tau}_{net}$ can also be calculated from:
\begin{equation}\label{tau2n}
\vec{\tau}_{net} = \int_{\Omega} \vec{r}\hspace{0.1cm}^{'}\times d\vec{F}^{ext}(\vec{r}) = \vec{\tau}_{cm}+\vec{R}_{cm}\times\vec{F}_{net},
\end{equation}
which by comparing to Eq. (\ref{tau1n}) allows us to conclude that, the torque about the $cm$ is given by $\vec{\tau}_{cm}=I_{cm}\vec{\alpha}$ and is valid regardless the kind of movement of the center of mass. The work done by the torque $\vec{\tau}_{cm}$ has the same form as that in Eq. (\ref{rotfan}) replacing $I_{O'}$ by $I_{cm}$. Finally, the total kinetic energy can be written as:  

\begin{equation}\label{n2n}
E_K=\frac{1}{2}\int_{\Omega} dm\,v'^{2} = \frac{1}{2}MV_{cm}^{2}+\frac{1}{2}I_{cm}\omega^{2}.
\end{equation}
From Eqs. (\ref{workcmn}) and (\ref{rotfan}) restricted to $O' \to cm$, it is possible to conclude that the contributions to the work related to the rotation and translation are separated because they are related to independent degrees of freedom. In contrast, for the case of rotation about a fixed axis, rotation and translation are associated and both contributions to the work are completely equivalent.

A more complete discussion of the results presented in this section can be found, for example, in Refs. \cite{kleppner,spiegel,meriam,lawden}.

\section{The disk and rod problem}

Now we apply the results of the previous section to analyze the role of the various contributions to the moment of inertia in the dynamic behavior of a system composed by several interacting rigid bodies. In order to accomplish this, we consider a simple physical system which have been widely studied: a compound physical pendulum \cite{kleppner,sol1,sol2,sol3,sol4}. Let's consider a disk of mass $M$ and radius $R$ fixed to the end of a rod of length $\ell$ and mass $m$, see Fig. \ref{dr} (a). In introductory courses of physics, the students are usually asked to determine the angular frequency in the limit of small oscillations. Two simple cases can be considered: in the first, the disk is fixed to the rod in such way that they form a single rigid body; in the second case, it is considered that the disk is mounted to the rod by a frictionless bearing, so that it is perfectly free to spin. 


\subsection{Case 1: Fixed Disk }
In the case 1, the rigid body formed by the rod and disk rotates around the fixed axis $O$. Then, from Eq. (\ref{n1n}) the angular momentum of the body with respect to the rotation axis is given by:
\begin{eqnarray}\label{lej1}
\vec{L}&=&\left(I_{rod}+I_{disk}\right)\vec{\omega}\nonumber\\
&=&\left(\frac{m\, \ell^2}{3}+M\,\ell^2+\frac{M\,R^2}{2}\right)\vec{\omega},
\end{eqnarray} 
where we have used the decomposition shown in Fig. \ref{cman} to calculate $I_{disk}$. This result is a particular case of the Steiner's theorem. 

On the other hand, taking the whole pendulum as a single rigid body, the torque with respect to the rotation axis due to the net force about $O$ depends only on the total weight of the body, see Fig.\ref{dr}-(b). The interaction between the disk and the rod is represented by the contact force $\vec{N}$ and the torque $\vec{\tau}$ as shown in Figs. \ref{dr}-(c) and (d). Note that the force $\vec{N}$ and the torque $\vec{\tau}$ are not zero, but they do not affect the motion of the rod-disk system. In this way, we have:
\begin{equation}\label{ejt1}
\vec{\tau}_{net}=-g\left(\frac{m\,\ell}{2}+M\,\ell\right)\sin\theta \hat{k}.
\end{equation}
Taking the time derivative on Eq. (\ref{lej1}) and matching the result with Eq. (\ref{ejt1}), it is easy to find the 
equation of motion for the pendulum:
\begin{equation}\label{ej1s}
\alpha=\frac{d^2\theta}{dt^2}=-\frac{g\,\ell\left( \frac{m}{2}+M\right) }{\frac{m\, \ell^2}{3}+M\,\ell^2+\frac{M\,R^2}{2}}\sin\theta. 
\end{equation}
This result can also be obtained easily by using Eq. (\ref{workcmn}) or (\ref{rotfan}). Taking Eq. (\ref{rotfan}) between the trajectory points defined by the angles $\theta$ and $\theta_0$, we have:
\begin{eqnarray}\label{ej1}
\mathcal{W}_{\theta}&=&-\int_{\theta_0}^{\theta} d\theta g\,\ell \left(\frac{m}{2}+M\right)\sin \theta\nonumber\\
&=&\frac{1}{2}I_0\left(\omega^2-\omega_0^2\right).
\end{eqnarray}
After integration, Eq. (\ref{ej1}) reduces to:
\begin{equation}\label{ej1a}
-g\,\ell\left(\frac{m}{2}+M \right)\cos \theta+\frac{1}{2} \left(\frac{m\, \ell^2}{3}+M\,\ell^2+\frac{M\,R^2}{2}\right)\omega^2=E,
\end{equation}
where $E$ is a constant which represents the total mechanical energy, more details can be found in Refs. \cite{sol1,sol2,sol3,sol4}. This is nothing more than the conservation energy equation. The first term represents gravitational potential energy, while the second the kinetic energy. Taking the time derivative of Eq. (\ref{ej1a}) we find:
\begin{equation}
\omega\left[g\,\ell\left(\frac{m}{2}+M \right)\sin \theta+\left(\frac{m\, \ell^2}{3}+M\,\ell^2+\frac{M\,R^2}{2}\right)\alpha\right]=0,
\end{equation}\label{ej1b}
where we have used $\alpha=d\omega/dt$ and $\omega=d\theta/dt$, arriving to Eq. (\ref{ej1s}).

\subsection{Case 2: Spinning disk}
In case 2, the disk can spin freely and the pendulum is not a rigid body any more. However, it is possible to describe the motion of the rod and disk separately as follows. From Eq. (\ref{n1n}), the angular momentum of the rod relative to the pivot $O'$ can be written as:
\begin{equation}\label{lej1a}
\vec{L}_{rod}=\frac{m\, \ell^2}{3}\vec{\omega}.
\end{equation}
As before, let $\vec{N}$ be the contact force between the disk and the rod. Nevertheless, now the disk can spin freely and, therefore, $\vec{\tau}=0$. Consequently, the net torque over the rod about $O$ is given by:
\begin{equation}\label{tej1}
\vec{\tau}_{net}=-\frac{m\,g\,\ell}{2}\sin\theta\,\hat{k}-\vec{\tau}_{N},
\end{equation}
where $\vec{\tau}_{N}$ is the torque applied by $\vec{N}$, see Fig. \ref{dr}-(c). From Eqs. (\ref{lej1a}) and (\ref{tej1}), the motion equation for the rod is:
\begin{equation}\label{ej1aa}
\frac{m\, \ell^2}{3}\alpha=-\frac{m\,g\,\ell}{2}\sin\theta-\tau_{N}.
\end{equation}

Now it is necessary to consider the motion of the disk. From the free body diagram it is clear that the torque about the $cm$ of the disk is null, see Fig \ref{dr} (d). Therefore, the disk rotates about its $cm$ with a constant angular velocity, let's say, $\vec{\beta}$. Therefore, the disk rotates and translates in such way that its angular momentum is given by Eq. (\ref{lrn}), leading to:
\begin{equation}\label{ldisk}
\vec{L}_{disk}=\frac{M\,R^2}{2} \vec{\beta}+M\,\ell^2\,\vec{\omega}.
\end{equation}  
The net torque about $O'$ can be written as:
\begin{equation}
\vec{\tau}_{net}=-M\,g\,\ell\sin\theta \, \hat{k} +\vec{\tau}_{N}.
\end{equation}
Consequently, the motion equation for the disk is:
\begin{equation}\label{ej1b}
M\,\ell^2 \alpha=-M\,g\,\ell\sin\theta+\tau_{N},
\end{equation}
where we have used the fact that $d\vec{\beta}/dt=0$. Note that Eq. (\ref{ej1b}) does not depend on $R$. The unknown torque $\tau_{N}$ can be eliminated by adding up Eqs. (\ref{ej1aa}) and (\ref{ej1b}), arriving to the desired motion equation:
\begin{equation}\label{ej1lr}
\alpha=-g\,\ell\frac{\frac{m}{2}+M}{\frac{m\,\ell^2}{3}+M\,\ell^2}\sin\theta. 
\end{equation}
As we can expect, Eq. (\ref{ej1lr}) can be also obtained applying Eq. (\ref{workcmn}) for the disk and Eq. (\ref{n1n}) for the rod. In this way, we have:
\begin{equation}\label{ejw1a}
\mathpzc{w}_{N} + M\,g\,\ell(\cos\theta-\cos\theta_0)=\frac{1}{2}M\,\ell^2\omega^2
\end{equation}
and
\begin{equation}\label{ejw1b}
-\mathpzc{w}_{N} + \frac{m\,g\,\ell}{2}(\cos\theta-\cos\theta_0)=\frac{1}{6}m\,\ell^2\omega^2,
\end{equation}
respectively. Adding up Eqs. (\ref{ejw1a}) and (\ref{ejw1b}) to eliminate the work done by the force $\vec{N}$, $\mathpzc{w}_N$, we arrive to the energy conservation equation:
\begin{equation}\label{pendener}
-\left( M\,g\,\ell+\frac{m\,g\,\ell}{2}\right) \cos\theta + \frac{1}{2} \left( M\,\ell^2 + \frac{m\,\ell^2}{3} \right) \omega^2=\text{const},
\end{equation}
where const is a constant, more details can be found in Refs. \cite{sol1,sol2,sol3,sol4}. Finally, taking the time derivative on Eq. (\ref{pendener}) as it was done on Eq. (\ref{ej1a}), we arrive to Eq. (\ref{ej1lr}).

Apart from what the students are usually asked to do in the limit of small oscillations $\left(\sin\theta \approx \theta\right)$ so that the harmonic solution takes place, a comparison between the frequencies of oscillation of the two situations can be made. Let $\vartheta$ be the oscillation frequency, therefore, for the Case 1,

\begin{equation}\label{freq1}
\vartheta^2_1 = g\ell\frac{\frac{m}{2}+M}{\frac{m\, \ell^2}{3}+M\,\ell^2+\frac{M\,R^2}{2}},
\end{equation}
and for Case 2, 
\begin{equation}\label{freq2}
\vartheta^2_2 = g\,\ell\frac{\frac{m}{2}+M}{\frac{m\,\ell^2}{3}+M\,\ell^2}.
\end{equation}

It is not hard to see that $\vartheta_1 < \vartheta_2$  regardless the values of the parameters. We conclude that, if the disk can spin freely around its $cm$, it will not contribute to any change of the overall angular momentum because it could be at most spinning at constant angular velocity, so the term related to its moment of inertia about its $cm$ will not appear in the overall moment of inertia. Therefore, a physical reason why the pendulum of case 1 takes longer to complete an oscillation is that it has larger moment of inertia, it is harder to make it rotate. Besides, we can imagine having to hold the rod from its free end and make whole pendulum rotate with our own hands. In the first case, the torque done is greater because we need to make the disk rotate exactly like the rod. In contrast, in the second case we do not need to make the disk rotate, it is free to spin, hence, if we apply the same torque of the previous case, it will oscillate faster 

It is important that the students note that, for both cases, the angular momentum of the disk has two contributions as shown in Eq. (\ref{ldisk}). The first term corresponds to rotation of the disk about its $cm$ while the second one to rotation of the $cm$ around the pivot. The time derivative of the first term is equal to $\vec{R}_{cm}\times \vec{F}_{net}$ which in turn depends on $\vec{N}$ and $\vec{W}_d$. The time derivative of the second term is equal to the internal torque applied by the rod, $\vec{\tau}$. For Case 1, $\vec{\tau}$ ensures that the disk rotates around the $cm$ with the angular velocity $\vec{\beta}=\vec{\omega}$, just as the rod does it about the pivot. However, in Case 2, $\vec{\tau}=0$ and the rotation of the disk about its $cm$ occurs at a constant angular velocity $\vec{\beta}$ which is not related to $\vec{\omega}$. An advance physics student would recognize the presence of a cyclic variable in the Hamiltonian and, consequently, the existence of a conserved quantity. Equivalently, a student of an introductory course should realize that for Case 2, $\vec{\tau}=0$, implying that the second term of the angular momentum given in Eq. (\ref{ldisk}) is a constant and does not contribute to the equation of motion. This result should be used to highlight to novice students the importance of conserved quantities.



%



\section{The free sliding ladder problem}


In this section we want to illustrate that whenever the mechanical energy is conserved, $\mathcal{W}_{\theta}$ and $\mathcal{W}_T$ may not vanish individually, though their sum must be zero. As in previous section, for illustrative purposes we use another problem widely used in introductory physics courses. Let's consider a ladder of length $2 \ell$ and mass $M$ which leans against a wall as shown in Fig. (\ref{ladder})-(a). All surfaces are frictionless. Due to the gravitational field, the ladder starts to slip downwards from the rest until eventually loses contact with the wall. In Ref. \cite{kleppner} the students are challenged to show that the ladder loses contact when it is at two-thirds of its initial height. In order to find this result, the students need to determine the angle for which the contact force with the wall becomes zero. However, for the purposes of our discussion we only need to calculate the rotational equation of motion of the ladder as follows. The position of the $cm$ can be written as $\vec{R}_{cm}= \ell \cos \theta  \hat{\imath} + \ell \sin \theta \hat{\jmath}$. On the other hand, the translational work due to the net force is given by:
\begin{eqnarray}\label{ej2a}
\mathcal{W}_T&=&\int_{\mathcal{C}} (\vec{N}_{w}+\vec{N}_{f}+\vec{W})\cdot d\vec{R}_{cm}\nonumber\\
 &=&\int_{\theta_0}^\theta d\theta\, \ell (N_{w}\hat{\imath}+(N_{f}-M\,g)\hat{\jmath})\cdot\left(-\sin \theta \hat{\imath}+\cos \theta \hat{\jmath}\right)\nonumber\\
&=&\int_{\theta_0}^\theta d\theta\, \ell (-N_{w} \sin \theta+N_{f}\cos \theta)-M\,g\, \ell(\sin \theta-\sin \theta_0)\nonumber\\
&=&\mathpzc{w}_c-M\,g\,\ell(\sin \theta-\sin \theta_0)\nonumber\\
&=&\frac{1}{2}M V_{cm}^2.
\end{eqnarray}
In Eq. (\ref{ej2a}), $\vec{N}_w$ and $\vec{N}_f$ are the contact forces applied by the wall and floor, respectively. Additionally, $\mathpzc{w}_c$ has been defined as the work due to these contact forces. Thus, from the last two lines of Eq. (\ref{ej2a}) we conclude:
\begin{equation}\label{ej2b}
\mathpzc{w}_c=\frac{1}{2}M V_{cm}^2+M g \ell\left(\sin \theta-\sin \theta_0\right).
\end{equation}
The net torque about the $cm$ is just due to $\vec{N}_w$ and $\vec{N}_f$. Therefore, by using Eq. (\ref{rotfan}), the rotational work associated to $\tau_{cm}$ takes the form:
\begin{equation}\label{ej2c}
\mathcal{W}_{\theta} =\int_{\theta_0}^\theta d\theta \,\ell (N_{w}\sin\theta-N_{f}\cos \theta)=-\mathpzc{w}_c.
\end{equation}
Then, replacing $I_{O}$ by $I_{cm}$ as discussed before Eq. (\ref{rotfan}), we find:
\begin{equation}\label{ej2ca}
-\mathpzc{w}_c=\frac{1}{6}M\,\ell^2\omega^2.
\end{equation}
Note that the rotational work of each contact force cancels out with its translational counterpart. Thus, the total mechanical energy is conserved despite $\mathcal{W}_{\theta}$ and $\mathcal{W}_T$ are both different to zero. In fact, adding Eqs. (\ref{ej2b}) and (\ref{ej2ca}) we arrive to:
\begin{equation}\label{ej2d}
\frac{1}{6}M\,\ell^2\omega^2+\frac{1}{2}M V_{cm}^2+M g\, \ell \sin \theta=E,
\end{equation}
where $E$ is a constant. This result has been previously derived in several texts and notes, for example, see Refs. \cite{spiegel,lawden,sol5}. As expected, Eq. (\ref{ej2d}) is the conservation energy equation. Finally, taking the time derivative of $\vec{R}_{cm}$ it is easy to find $V_{cm}=\ell\,\omega$. Inserting this result in Eq. (\ref{ej2d}) we find:
\begin{equation}\label{ej2e}
\frac{2}{3} M \ell^2 \omega^2+M g\,\ell \sin \theta=E.
\end{equation}
As before, from the time derivative of Ec. (\ref{ej2e}) we arrive to the motion equation in terms of the variable $\theta$:
\begin{equation}\label{ej2f}
\alpha=-\frac{3\,g}{4\,\ell}\cos \theta.
\end{equation}

In many particle systems, whether they are rigid or not, it is always possible to describe the motion of any particle of the system through the motion of the $cm$ and the relative motion of the particle about the $cm$. This decomposition allows us to write the theorem of work and energy in two separate parts: translation ($\mathcal{W}_T$) and rotation ($\mathcal{W}_{\theta}$). The former is  the work done by the net force along the path of the $cm$. The latter is the work done by the net torque with respect to the $cm$. In the simple case where the body rotates around a fixed axis, $\mathcal{W}_{\theta}=\mathcal{W}_T$ and both parts of the theorem of work and energy  are completely equivalent. However, in the most general case which involves rotation about the $cm$ and translation of the $cm$,  $\mathcal{W}_T$ and  $\mathcal{W}_{\theta}$, are not necessarily equal. In fact, for systems where the total mechanical energy is conserved, it is satisfied that the sum of the rotational and translational works associated to non-conservative forces is zero.

\begin{figure}[htp]
\begin{center}
\includegraphics[scale=0.4]{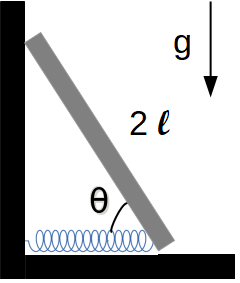}
\end{center}
\caption{Sliding ladder with a spring.}
\label{ladderSpring}
\end{figure}
On the other hand, a deeper observation of Eqs. (\ref{ej2b}) and (\ref{ej2c}) shows that the contact forces are completely defined by the geometrical constraint:
\begin{equation}\label{cons}
x_{cm}^2+y_{cm}^2=\ell^2,
\end{equation}
in such way that the total work associated to the contact forces is zero regardless the value of these forces. Naturally, an advanced physics student would easily realize that this a consequence of the holonomic nature of the constraint (\ref{cons}). The total work of the forces associated to holonomic constraints, as the one given by Eq. (\ref{cons}), is always zero \cite{gold,greiner}. However, this is not true for all kind of forces. In order to illustrate this idea consider the sliding ladder with an spring as shown in Fig. (\ref{ladderSpring}). In this case, the force applied by the spring is $\vec{F}_{s} = -k \Delta x\,\hat{i}$, where $\Delta x = 2\ell\left(\cos \theta - \cos \theta_{0}\right)$. Therefore, the rotational and translational works associated to the force applied by the spring, $w_{\theta}$ and $w$, satisfy:
\begin{equation}
\mathpzc{w}_{\theta}=\mathpzc{w}_T=\int_{\theta_0}^\theta k(2\ell \cos \theta-l_0)\ell\,\sin \theta\, d\theta.
\end{equation}
Assuming that the system is released when the spring has its natural length ($l_0=2\,\ell\,\cos \theta_0$), we find:
\begin{equation}
\mathpzc{w}_{\theta}=\mathpzc{w}_T=-k\ell^2(\cos \theta-\cos \theta_0)^{2},
\end{equation}
in such way that $\mathpzc{w}_{\theta}+\mathpzc{w}_T$ is minus the potential energy associated to the spring.



\section{Discussion}

As mentioned in the introduction, the solution of the compound pendulum problem can be found in several informal physics forums. However, in spite that those solutions show the correct final result, they usually lack of a rigorous explanation of the main conclusion: $\vartheta_1<\vartheta_2$ given that, in Case 2, the moment of inertia of the disk relative to its $cm$ does not contribute. In fact, sentences such as  ``The term $M\,R^2/2$ should then be omitted from...''  and  ``If the disk were free so that it doesn’t rotate at all, it’s not hard to convince yourself that it acts just like a point mass (because it doesn’t rotate as the rod swings)", among others, are included to support the final result \cite{p1,sol1,sol2,sol3}. In our experience, many students intuitively understand why $\vartheta_2>\vartheta_1$ and accept the result without the need of a formal derivation. However, we strongly suggest to present in introductory physics courses a complete derivation, as the one provided in this paper, including an explanation based on the description interaction between the rod and disk ($\vec{\tau}$ and $\vec{N}$) and on basic physical laws. Otherwise, we run the risk to accentuate the erroneous idea that, in Case 2, the moment of inertia of the disk becomes zero because it can rotate freely.

On the other hand, from Eqs. (\ref{wsum}), (\ref{workcmn}), (\ref{rotfan}) and (\ref{n2n}) it is clear that:
\begin{equation}\label{wrel}
\mathcal{W}=\mathcal{W}_T+\mathcal{W}_{\theta}=\Delta E_k,
\end{equation}
where the $\mathcal{W}$ is the total work due the external forces over the body given by Eq. (\ref{wsum}). On the other hand, the total work is the sum of the work due to conservative and non-conservative forces, $\mathcal{W}_c$ and $\mathcal{W}_{nc}$, respectively. As usual, $\mathcal{W}_c$ can be written as the change of the total potential energy, $-\Delta E_p$. Thus, it is possible to conclude that:
\begin{equation}
\mathcal{W}_T+\mathcal{W}_{\theta}=\mathcal{W}_{nc}-\Delta E_p.
\end{equation}
Now, $\mathcal{W}_T$ and $\mathcal{W}_{\theta}$ can be also decomposed in the sum of the contributions of conservative and non-conservative forces. Thus, we write:
\begin{equation}
\mathcal{W}^c_T+\mathcal{W}^{nc}_T+\mathcal{W}^c_{\theta}+\mathcal{W}^{nc}_{\theta}=\mathcal{W}_{nc}-\Delta E_p.
\end{equation}
If the total mechanical energy of the body is conserved the total work due the non-conservative forces satisfy $\mathcal{W}_{nc}=0$. Given that the potential energy depends on the parameters that defines the conservative forces such as the spring constant or the gravity constant, the last equation implies that:
\begin{equation}
\mathcal{W}^{nc}_T+\mathcal{W}^{nc}_{\theta}=0\,\,\,\text{and}\,\,\,\mathcal{W}^c_T+\mathcal{W}^c_{\theta}=-\Delta E_p.
\end{equation}
Therefore, it is possible to conclude that, as usual, for conservative systems the non-conservative forces do not contribute to the total work $\mathcal{W}$ but they could perform translational and rotational works as long as the sum of them vanish.
In introductory courses, the separation of the work into two parts improves the understanding of the students about the RBP motion. For instance, this approach allows students to realize that, in many cases, non-conservative forces could do translational and rotational work despite the total energy of the system is conserved. The ladder problem is an example of that. Usually, for the sake of simplicity, in many cases solutions to the problems solved in class use directly the energy conservation equation. This equation usually is easy to use and simplifies the solution of many problems. However, it can hide some important information about the system. For instance, the existence of non-conservative forces which perform translational and rotational works on a system where the mechanical energy is conserved. Additionally, the conservation energy approach usually puts in a second place some concepts which are equally important such as the work and energy theorem. This theorem involves an integral, and because of that, for the students it is more complicated to understand. However, the use of this theorem helps to clarify physical situations as the one mentioned here. We recommend to find the solution of the problems using different methods such as the angular momentum/torque, work and energy theorem, energy conservation, etc. This makes easier for students the understanding of the relation among different physical concepts, for example, helping them to understand that the presence of non-conservative forces does not necessarily imply that energy is not conserved.
	%
	
	
	
	
	
	
	
	
	

\section*{Acknowledgments}
The work of D.L.G was supported by the Vicerrectoría de investigaciones de la Universidad del Valle C.I. 1164. Y.R. was supported by the following grants: Colciencias-Deutscher Akademischer Austauschdienst Grant No. 110278258747 RC-774-2017, Vicerrectoría de Ciencia, Tecnología, e Innovación - Universidad Antonio Nariño Grant No. 2019248, and Dirección de Investigación y Extensión de la Facultad de Ciencias - Universidad Industrial de Santander Grant No. 2460.


\end{document}